\documentclass[a4paper,12pt]{article}
\usepackage{amssymb}
\usepackage{amsmath}
\usepackage{epsfig}

\usepackage{graphics}

\usepackage{latexsym}
\usepackage{rotating}
\title{ Nonlocal Fordy - Kulish  Equations on Symmetric Spaces}

\author {
Metin G{\" u}rses \thanks{Email:gurses@fen.bilkent.edu.tr}\\
{\small Department of Mathematics, Faculty of Science} \\
{\small Bilkent University, 06800 Ankara - Turkey}}

\date{\nonumber}
\begin{document}
\maketitle
\date{\nonumber}

\baselineskip 17pt

\numberwithin{equation}{section}

\begin{abstract}
We present nonlocal integrable reductions of the Fordy-Kulish system of nonlinear Schrodinger equations and the Fordy system
of derivative nonlinear Schrodinger equations on Hermitian symmetric spaces. Examples are given on the symmetric space $\frac{SU(4)}{SU(2) \times SU(2)}$.
\end{abstract}

\newtheorem{thm}{Theorem}[section]
\newtheorem{Le}{Lemma}[section]
\newtheorem{defi}{Definition}[section]

\def \part {\partial}
\def \be {\begin{equation}}
\def \ee {\end{equation}}
\def \bea {\begin{eqnarray}}
\def \eea {\end{eqnarray}}
\def \ba {\begin{array}}
\def \ea {\end{array}}
\def \si {\sigma}
\def \al {\alpha}
\def \la {\lambda}
\def \D {\displaystyle}


\section{Introduction}
Recently Ablowitz and Musslimani \cite{ab1}-\cite{ab3} have given a new reduction of coupled system of integrable nonlinear equations. This
reduction is based on the time (T), space (P) and space-time (PT) reflections of the half of the dynamical systems preserving the
integrability property of the system. As an example we can give the integrable coupled AKNS system

\begin{eqnarray}
a\,q_{t}&=& q_{xx}-2 q^2\, p,\label{denk1}\\
a\,p_{t}&=&-p_{xx}+2 p^2\, q. \label{denk2}
\end{eqnarray}
where $p(t,x)$ and $q(t,x)$ are complex dynamical variables and $a$ is a complex number in general. The standard reduction of this is system
is obtained by letting $p(t,x)=\epsilon q(t,x)^{\star}$ where $\star$ is the complex conjugation. With this condition on the dynamical variables $q$ and $p$ the system
of equations (\ref{denk1}) and (\ref{denk2}) reduce to the following nonlinear Schrodinger equation (NLS)

\begin{equation}
a\,q_{t}=q_{xx}-2 \epsilon q^2\, q^{\star}, \label{denk3}
\end{equation}
provided that $a^{\star}=-a$. Recently Ablowitz and Musslimani found another integrable reduction which we call it as a nonlocal reduction of the above AKNS system (\ref{denk1}) and (\ref{denk2}) which is given by

\begin{equation}
p(t,x)=\epsilon [q(\epsilon_{1} t, \epsilon_{2} x)]^{\star} \label{non}
\end{equation}
where $ \epsilon^2=(\epsilon_{1})^2=(\epsilon_{2})^2=1$. Under this
condition the above AKNS system (\ref{denk1}) and (\ref{denk2}) reduce to

\begin{equation}
a \,q_{t,x}=q_{xx}(t,x) -2\epsilon  q^2(t,x)\, [q(\epsilon_{1} t, \epsilon_{2} x)]^{\star}, \label{denk4}
\end{equation}
provided that $a^{\star}=- \epsilon_{1} a$. Nonlocal reductions correspond to $(\epsilon_{1}, \epsilon_{2})=(-1,1),(1,-1),(-1,-1)$. Hence for these values of $\epsilon_{1}$ and  $\epsilon_{2}$ we have six different Nonlocal Integrable NLS equations. They are respectively the time reflection symmetric (T-symmetric), the space reflection symmetric (P-symmetric) and the space-time reflection symmetric (PT-symmetric) nonlocal nonlinear Schrodinger equations which are given by

\vspace{0.3cm}
\noindent
1. T-symmetric nonlinear Schrodinger equation

\begin{equation}
a \,q_{t,x}=q_{xx}(t,x) -2 \epsilon  q^2(t,x)\, [q(- t, x)]^{\star}, ~~~a^{\star}=a.
\end{equation}

\vspace{0.3cm}
\noindent
2. P-symmetric nonlinear Schrodinger equation

\begin{equation}
a \,q_{t,x}=q_{xx}(t,x)-2 \epsilon  q^2(t,x)\, [q(t, -x)]^{\star},~~~a^{\star}=-a.
\end{equation}

\vspace{0.3cm}
\noindent
3. PT-symmetric nonlinear Schrodinger equation

\begin{equation}
a \,q_{t,x}=q_{xx}(t,x) -2 \epsilon  q^2(t,x)\, [q(-t, -x)]^{\star},~~~a^{\star}=a.
\end{equation}

Ablowitz and Musslimani have found many other nonlocal integrable equations such as nonlocal modified KdV equation, nonlocal Davey-Stewartson equation, nonlocal sine-Gordon equation and nonlocal $2+1$ dimensional three-wave interaction equations \cite{ab1}-\cite{ab3}. There is a recent increasing interest on the Ablowitz-Musslimani nonlocal reductions of system of coupled integrable nonlinear equations. For instance Fokas \cite{fok} has introduced nonlocal versions of multidimensional Schrodinger equation, Ma et al., \cite{ma} pointed out that nonlocal complex mKdV equation introduced by Ablowitz and Musslimani is gauge equivalent to a spin-like model, Gerdjikov and Saxena \cite{gerd} studied the complete integrability of nonlocal nonlinear Schrodinger
equation and Sinha and Ghosh \cite{sin} have introduced the nonlocal vector nonlinear Schrodinger equation.

In this letter we give all possible nonlocal reductions of the Fordy-Kulish system \cite{fk}  of coupled nonlinear Schrodinger and Fordy system \cite{for1}  of coupled derivative nonlinear Schrodinger equations on Hermitian symmetric spaces. Nonlocal system equations were also studied recently by Gerdjikov, Grahovski and Ivanov \cite{gerd1}- \cite{gerd3}.

Although we will study the 1-Soliton solutions of the nonlocal Fordy-Kulish and nonlocal Fordy equations in a future communication but we would like
to make a comment on an important feature of the recently found 1-soliton solutions of some nonlocal equations.
It was noticed that 1-soliton solutions of the nonlocal NLSE develope  singularities in a finite time \cite{ab1}. It is highly probable  that such a singular structure may arise also in 1-soliton solutions of the system of nonlocal Fordy-Kulish equations. On the other hand it was observed that multi-component derivative NLSE can have regular 1-soliton solutions \cite{gerd1}. In this respect we expect that nonlocal system of nonlocal Fordy equations can have also regular 1-soliton solutions.

\section{Fordy-Kulish System}

Systems of integrable nonlinear partial differential equations arise when the Lax pairs are given in certain Lie algebras. Fordy-Kulish and Fordy systems of equations are examples of such equations. We briefly give the Lax representations of these equations. For more details see \cite{fk}, \cite{for1}, \cite{gerd1}, \cite{gur1}.

Let $G$ be a Lie group. A homogeneous space of $G$ is any differentiable manifold $M$ on which $G$ acts transitively. If $K$ is an isotropy subgroup of $G$, $M$ is identified with a coset space $G/K$. Let ${\it{g}}$ and ${\it{k}}$ be the Lie algebras of $G$ and $K$ respectively. Let ${\it m}$  be the vector space complement of $\it{k}$ in $\it{g}$. Then $\it{g}=\it{k} \oplus \it{m}$ with $[\it{k}, \it{k}] \in \it{k}$. If, furthermore $[\it{k}, \it{m}] \in \it{m}$ and
$[\it{m}, \it{m}] \in \it{k}$ then $G/K$ is called symmetric space. $H_{a}$'s are commuting generators of $G$ with $a=1,2.\cdots, P$ where $P$ is the rank of the Lie algebra. $H_{a}$ and $E_{D}$ are the generators of the subgroup $K$.  Consider the linear equations (Lax equations)

\begin{eqnarray}
\phi_{x}=(\lambda H_{S}+Q^{A}\, E_{A})\, \phi,  \label{lax1} \\
\phi_{t}=(A^a H_{a}+B^{A} E_{A}+C^{D} E_{D})\, \phi \label{lax2}
\end{eqnarray}
where the dynamical variables are $Q_{A}=(q^{\alpha},p^{\alpha})$, the functions $A^{a}$, $B^{A}$ and $C^{A}$ depend on the spectral parameter $\lambda$, on the dynamical variables ($q^{\alpha} $, $p^{\alpha}$) and their $x$-derivatives.
The integrability condition $\phi_{xt}=\phi_{tx}$ leads to systems of nonlinear partial differential equations of the evolutionary type. The system of Fordy-Kulish equations is an example when the functions $A$, $B$ and $C$ are quadratic functions of $\lambda$.
Let $q^{\alpha}(t,x)$ and $p^{\alpha}(t,x)$ be the complex dynamical variables where $\alpha=1,2,\cdots,N$, then the Fordy-Kulish (FK) integrable system arising from the integrability condition of Lax equations (\ref{lax1}) and (\ref{lax2})
is given by \cite{fk}

\begin{eqnarray}
a q^{\alpha}_{t}&=& q^{\alpha}_{xx}+ R^{\alpha}\,_{\beta \gamma -\delta}\, q^{\beta}\,q^{\gamma}\, p^{\delta}, \label{denk5},\\
-a p^{\alpha}_{t}&=& p^{\alpha}_{xx}+ R^{-\alpha}\,_{-\beta -\gamma \delta}\, p^{\beta}\,p^{\gamma}\, q^{\delta}, \label{denk6},
\end{eqnarray}
for all $\alpha=1,2,\cdots,N$ where $R^{\alpha}\,_{\beta \gamma -\delta},R^{-\alpha}\,_{-\beta -\gamma \delta}$ are the curvature tensors of a Hermitian symmetric space satisfying
\begin{equation}
(R^{\alpha}\,_{\beta \gamma -\delta})^{\star}=R^{-\alpha}\,_{-\beta -\gamma \delta},\label{prop}
\end{equation}
and $a$ is a complex number. Here we use the summation convention, i.e., the repeated indices are summed up from 1 to $N$. These equations are known as the Fordy-Kulish (FK) system which is integrable in the sense that they are obtained from the zero curvature condition of a connection defined on a Hermitian symmetric space and these equations can also be written in a Hamiltonian form

\begin{equation}
a q^{\alpha}_{t}=g^{\alpha -\beta}\, \frac{\delta H}{\delta p^{\beta}},~~ -a p^{\alpha}_{t}=g^{\alpha -\beta}\, \frac{\delta H}{\delta q^{\beta}}
\end{equation}
where
\begin{equation}
H=-g_{\alpha -\beta}\,q^{\alpha}_{x}\, p^{\beta}_{x}+\frac{1}{2}\, g_{\epsilon -\alpha}\, R^{\epsilon}\,_{\beta \gamma -\delta}\, p^{\alpha}\,q^{\beta}\,q^{\gamma}\, p^{\delta}
\end{equation}
where $g_{\alpha-\beta}$ are the components of the metric tensor
The standard reduction of the above FK is obtained by letting $p^{\alpha}=\epsilon (q^{\alpha})^{\star}$ for all $\alpha=1,2,\cdots,N$ and $\epsilon^2=1$. The FK system (\ref{denk5}) and (\ref{denk6}) reduce to a single equation

\begin{equation}
a q^{\alpha}_{t}=q^{\alpha}_{xx} +\epsilon \, R^{\alpha}\,_{\beta \gamma -\delta}\, q^{\beta}\,q^{\gamma}\,(q^{\delta})^{\star}, ~~ \alpha=1,2,\cdots,N \label{denk7},
\end{equation}
provided that $a^{\star}=-a$ and (\ref{prop}) is satisfied.

\section{Nonlocal Fordy-Kulish System}

Here we will show that the Fordy-Kulish system is compatible with the nonlocal reduction of Ablowitz-Musslimani type. For this purpose
using a similar condition as (\ref{non}) we let

\begin{equation}
p^{\alpha}(t,x)=\epsilon [q^{\alpha}(\epsilon_{1} t, \epsilon_{2} x)]^{\star}, ~~~\alpha=1,2,\cdots,N
\end{equation}
where $ \epsilon^2=(\epsilon_{1})^2=(\epsilon_{2})^2=1$. Under this constraint the Fordy-Kulish system (\ref{denk5}) and (\ref{denk6})
reduce to the following system of equations.

\begin{equation}
a q^{\alpha}_{t}(t,x)=q^{\alpha}_{xx}(t,x) +\epsilon \, R^{\alpha}\,_{\beta \gamma -\delta}\, q^{\beta}(t,x)\,q^{\gamma}(t,x)\,(q^{\delta}(\epsilon_{1} t, \epsilon_{2} x))^{\star},  \label{denk9},
\end{equation}
provided that $a^{\star}=-\epsilon_{1}\,a$ and (\ref{prop}) is satisfied. In addition to (\ref{denk9}) we have also equation for $q^{\delta}(\epsilon_{1} t, \epsilon_{2} x)$ which can be obtained by letting $t \rightarrow \epsilon_{1}t$, $x \rightarrow \epsilon_{2}x$ in (\ref{denk9}). Hence we obtain $T-$ symmetric, $P-$ symmetric and $PT-$ symmetric nonlocal nonlinear Schrodinger equations.  Nonlocal reductions correspond to $(\epsilon_{1}, \epsilon_{2})=(-1,1),(1,-1),(-1,-1)$. Hence corresponding to these values of $\epsilon_{1}$ and  $\epsilon_{2}$ we have six different Nonlocal Integrable NLS equations. They are given as follows:

\vspace{0.3cm}
\noindent
1. T-symmetric nonlocal FK system:

\begin{equation}
a q^{\alpha}_{t}(t,x)=q^{\alpha}_{xx}(t,x)+ \epsilon \, R^{\alpha}\,_{\beta \gamma -\delta}\, q^{\beta}(t,x)\,q^{\gamma}(t,x)\,(q^{\delta}(-t,x))^{\star},  \label{denk10},
\end{equation}
with $a^{\star}=a$.

\vspace{0.3cm}
\noindent
2. P-symmetric nonlocal FK system:

\begin{equation}
a q^{\alpha}_{t}(t,x)=q^{\alpha}_{xx}(t,x)+ \epsilon \, R^{\alpha}\,_{\beta \gamma -\delta}\, q^{\beta}(t,x)\,q^{\gamma}(t,x)\,(q^{\delta}(t,-x))^{\star},  \label{denk11},
\end{equation}
with $a^{\star}=-a$.

\vspace{0.3cm}
\noindent
3. PT-symmetric nonlocal FK system:

\begin{equation}
a q^{\alpha}_{t}(t,x)=q^{\alpha}_{xx}(t,x) +\epsilon \, R^{\alpha}\,_{\beta \gamma -\delta}\, q^{\beta}(t,x)\,q^{\gamma}(t,x)\,(q^{\delta}(-t,-x))^{\star},  \label{denk11}
\end{equation}
with $a^{\star}=a$. All these six nonlocal equations are integrable.

\vspace{0.3cm}
\noindent
{\bf Example 1}: An explicit example for nonlocal FK system is on the symmetric space  $\frac{SU(4)}{SU(2) \times SU(2)}$.
In this case we have four complex fields $q^{\alpha}=(q^1(t,x),q^2(t,x),q^3(t,x),q^4(t,x))$ satisfying the following equations (we take $\epsilon=1$)

\vspace{0.3cm}
\noindent
1. T-symmetric nonlocal FK system:

\begin{eqnarray}
a q^1_{t}(t,x)&=&q^{1}_{xx}(t,x)+2 q^{1}(t,x)[q^{1}(t,x) (q^{1}(-t,x))^{\star}+q^{2}(t,x) (q^{2}(-t,x))^{\star}+q^{4}(t,x) (q^{4}(-t,x))^{\star}] \nonumber\\
&&+2 q^{2}(t,x)\,q^{4}(t,x)\,(q^{3}(-t,x))^{\star}, \\
a q^2_{t}(t,x)&=&q^{2}_{xx}(t,x)+2 q^{2}(t,x)[q^{1}(t,x) (q^{1}(-t,x))^{\star}+q^{2}(t,x) (q^{2}(-t,x))^{\star}+q^{3}(t,x) (q^{3}(-t,x))^{\star}] \nonumber \\
&&+2 q^{1}(t,x)\,q^{3}(t,x)\,(q^{4}(-t,x))^{\star}
\end{eqnarray}
with $a^{\star}=a$. The remaining two equations are obtained by using the above equations interchanging $q^{1} \leftrightarrow q^{3}$ and $q^{2} \leftrightarrow q^{4}$.

\vspace{0.3cm}
\noindent
2. P-symmetric nonlocal FK system:

\begin{eqnarray}
a q^1_{t}(t,x)&=&q^{1}_{xx}(t,x)+2 q^{1}(t,x)[q^{1}(t,x) (q^{1}(t,-x))^{\star}+q^{2}(t,x) (q^{2}(t,-x))^{\star}+q^{4}(t,x) (q^{4}(t,-x))^{\star}] \nonumber\\
&&+2 q^{2}(t,x)\,q^{4}(t,x)\,(q^{3}(-t,x))^{\star}, \\
a q^2_{t}(t,x)&=&q^{2}_{xx}(t,x)+2 q^{2}(t,x)[q^{1}(t,x) (q^{1}(t,-x))^{\star}+q^{2}(t,x) (q^{2}(t,-x))^{\star}+q^{3}(t,x) (q^{3}(t,-x))^{\star}] \nonumber \\
&&+2 q^{1}(t,x)\,q^{3}(t,x)\,(q^{4}(t,-x))^{\star}
\end{eqnarray}
with $a^{\star}=-a$. The remaining two equations are obtained by using the above equations interchanging $q^{1} \leftrightarrow q^{3}$ and $q^{2} \leftrightarrow q^{4}$.

\vspace{0.3cm}
\noindent
3. PT-symmetric nonlocal FK system:

\begin{eqnarray}
a q^1_{t}(t,x)&=&q^{1}_{xx}(t,x)+2 q^{1}(t,x)[q^{1}(t,x) (q^{1}(-t,-x))^{\star}+q^{2}(t,x) (q^{2}(-t,-x))^{\star}+q^{4}(t,x) (q^{4}(-t,-x))^{\star}] \nonumber\\
&&+2 q^{2}(t,x)\,q^{4}(t,x)\,(q^{3}(-t,-x))^{\star}, \\
a q^2_{t}(t,x)&=&q^{2}_{xx}(t,x)+2 q^{2}(t,x)[q^{1}(t,x) (q^{1}(-t,-x))^{\star}+q^{2}(t,x) (q^{2}(-t,-x))^{\star}+q^{3}(t,x) (q^{3}(-t,-x))^{\star}] \nonumber \\
&&+2 q^{1}(t,x)\,q^{3}(t,x)\,(q^{4}(-t,-x))^{\star}
\end{eqnarray}
with $a^{\star}=a$. The remaining two equations are obtained by using the above equations interchanging $q^{1} \leftrightarrow q^{3}$ and $q^{2} \leftrightarrow q^{4}$.

\section{Fordy System}

Fordy, in his work \cite{for1} considering the Lax equations

\begin{eqnarray}
\phi_{x}=( \lambda^2\, H_{S}+\lambda\, Q^{A}\, E_{A})\, \phi,  \label{lax3} \\
\phi_{t}=(A^a H_{a}+B^{A} E_{A}+C^{D} E_{D})\, \phi \label{lax4}
\end{eqnarray}
and choosing functions $A^{a}$, $ B^{A}$, and $C^{A}$ properly
he has found system of derivative nonlinear Schrodinger equations on homogenous spaces. Here we use this system on Hermitian symmetric spaces. Let $Q^{A}=\{q^{\alpha}(t,x), p^{\alpha}(t,x)\}$ be the complex dynamical variables where $\alpha=1,2,\cdots,N$, then the Fordy (F) integrable system is given by \cite{for1}

\begin{eqnarray}
a q^{\alpha}_{t}&=& [q^{\alpha}_{x}-\frac{1}{a}\,R^{\alpha}\,_{\beta \gamma -\delta}\, q^{\beta}\,q^{\gamma}\, p^{\delta}]_{x}, \label{denk12},\\
-a p^{\alpha}_{t}&=& [p^{\alpha}_{x}+ \frac{1}{a}\,R^{-\alpha}\,_{-\beta -\gamma \delta}\, p^{\beta}\,p^{\gamma}\, q^{\delta}]_{x}, \label{denk13},
\end{eqnarray}
where $a$ is a complex number. These equations are known as the Fordy (F) system which is integrable in the sense that they are obtained from the zero curvature condition of a connection defined on a Hermitian symmetric space and also these equations can be written in a Hamiltonian form

\begin{equation}
a q^{\alpha}_{t}=g^{\alpha -\beta}\, \partial \,\frac{\delta H}{\delta p^{\beta}},~~a p^{\alpha}_{t}=g^{\alpha -\beta}\, \partial \frac{\delta H}{\delta q^{\beta}}
\end{equation}
where
\begin{equation}
H=-\frac{1}{2}[g_{\alpha -\beta}\,(q^{\alpha}\, p^{\beta}_{x}-q^{\alpha}_{x}\, p^{\beta})+\frac{1}{a}\, g_{\rho -\alpha}\, R^{\rho}\,_{\beta \gamma -\delta}\, p^{\alpha}\,q^{\beta}\,q^{\gamma}\, p^{\delta}]
\end{equation}
The standard reduction of the above F system is obtained by letting $p^{\alpha}=\epsilon (q^{\alpha})^{\star}$ for all $\alpha=1,2,\cdots,N$. The F system (\ref{denk12}) and (\ref{denk13}) reduce to a single equation

\begin{equation}
a q^{\alpha}_{t}=[q^{\alpha}_{x} -\frac{\epsilon}{a} \, R^{\alpha}\,_{\beta \gamma -\delta}\, q^{\beta}\,q^{\gamma}\,(q^{\delta})^{\star}]_{x},  \label{denk7},
\end{equation}
provided that $a^{\star}=-a$ and (\ref{prop}) is satisfied. This is the coupled derivative nonlinear Schrodinger equations.

\section{Nonlocal Fordy System}

We will show that the F-system is also compatible with for nonlocal reduction of Ablowitz-Musslimani type. Letting

\begin{equation}
p^{\alpha}(t,x)=\epsilon [q^{\alpha}(\epsilon_{1} t, \epsilon_{2} x)]^{\star}, ~~~\alpha=1,2,\cdots,N
\end{equation}
where $ \epsilon^2=(\epsilon_{1})^2=(\epsilon_{2})^2=1$. Under these conditions the Fordy system (\ref{denk12}) and (\ref{denk13})
reduce to the following system of equations.

\begin{equation}
a q^{\alpha}_{t}(t,x)=[q^{\alpha}_{x}(t,x) -\epsilon \frac{1}{a}\, R^{\alpha}\,_{\beta \gamma -\delta}\, q^{\beta}(t,x)\,q^{\gamma}(t,x)\,(q^{\delta}(\epsilon_{1} t, \epsilon_{2} x))^{\star}]_{x},  \label{denk14},
\end{equation}
provided that $a^{\star}=-\epsilon_{1}\,a$, $\epsilon_{1}\,\epsilon_{2}=1$ and (\ref{prop}) satisfied. Nonlocal reductions correspond to $(\epsilon_{1}, \epsilon_{2})=(-1,-1)$. Hence corresponding to these pairs of $\epsilon_{1}$ and  $\epsilon_{2}$ we have two different Nonlocal Integrable derivative NLS equations. They are given as follows:

\vspace{0.3cm}
\noindent
PT-symmetric nonlocal F system:

\begin{equation}
a q^{\alpha}_{t}(t,x)=[q^{\alpha}_{x}(t,x) -\epsilon \frac{1}{a}\, R^{\alpha}\,_{\beta \gamma -\delta}\, q^{\beta}(t,x)\,q^{\gamma}(t,x)\,(q^{\delta}(-t,-x))^{\star}]_{x},  \label{denk116},
\end{equation}
with $a^{\star}=a$.

\vspace{0.3cm}
\noindent
{\bf Example 2}: As in the case of the FK system we give an example of F system on the symmetric space as $\frac{SU(4)}{SU(2) \times SU(2)}$. We have four complex fields $q^{\alpha}=(q^1(t,x),q^2(t,x),q^3(t,x),q^4(t,x))$ satisfying the following equations (here we take $\epsilon=1$)

\vspace{0.3cm}
\noindent
PT-symmetric nonlocal F system:

\begin{eqnarray}
a q^1_{t}(t,x)&=&(q^{1}_{x}(t,x)- \frac{2}{a} q^{1}(t,x)[q^{1}(t,x) (q^{1}(-t,-x))^{\star}+q^{2}(t,x) (q^{2}(-t,-x))^{\star}+q^{4}(t,x) (q^{4}(-t,-x))^{\star}] .\nonumber\\
&&-\frac{2}{a} q^{2}(t,x)\,q^{4}(t,x)\,(q^{3}(-t,-x))^{\star})_{x}, \\
a q^2_{t}(t,x)&=&(q^{2}_{x}(t,x)-\frac{2}{a} q^{2}(t,x)[q^{1}(t,x) (q^{1}(-t,-x))^{\star}+q^{2}(t,x) (q^{2}(-t,-x))^{\star}+q^{3}(t,x) (q^{3}(-t,-x))^{\star}] \nonumber \\
&&-\frac{2}{a} q^{1}(t,x)\,q^{3}(t,x)\,(q^{4}(-t,-x))^{\star})_{x}
\end{eqnarray}
with $a^{\star}=a$. The remaining two equations are obtained by using the above equations interchanging $q^{1} \leftrightarrow q^{3}$ and
$q^{2} \leftrightarrow q^{4}$.

\section{Concluding Remarks}

By using the Ablowitz-Musslimani type of reductions on dynamical systems we have shown that the Fordy-Kulish and Fordy systems of equations on Hermitian symmetric spaces we obtained to nonlocal integrable dynamical equations possessing T-, P-, and PT-symmetries. Such a reduction can be extended to system of integrable equations on homogeneous spaces of simple Lie algebras, Kac-Moody algebras and super Lie algebras.
From the Lax pair (\ref{lax1}) and (\ref{lax2}) we can construct  a Lie algebra ${\it g}$ valued soliton connection
\begin{equation}
\Omega=(i k \lambda H_{S}+Q^{A}\, E_{A}) dx+(A^a H_{a}+B^{A} E_{A}+C^{D} E_{D}) dt \label{con}
\end{equation}
with zero curvature $d \Omega-\Omega \wedge \Omega=0$.  In the soliton connection the dynamical variables are $Q_{A}=(q^{\alpha},p^{\alpha})$. The functions $A^{a}$, $B^{A}$ and $C^{A}$ depend on the spectral parameter $\lambda$, the dynamical variables ($q^{\alpha} $, $p^{\alpha}$) and their $x$-derivatives. The zero curvature condition for (\ref{con}) gives systems coupled nonlinear partial differential equations of the evolutionary type. The system of Fordy-Kulish equations is an example when the functions $A$, $B$ and $C$ are quadratic functions of $\lambda$. One can extend this approach to the cases when ${\it g}$ is a Koc-Moody  and super Lie algebras (see \cite{gur1} for more details). For all the system of dynamical equations we obtain from the ${\it g}$-valued soliton connection $\Omega$ we conjecture that Ablowitz-Musslimani type constraint for the dynamical variables $p$ and $q$, namely
\begin{equation}
p^{\alpha}=\epsilon [q^{\alpha}(\epsilon_{1} t, \epsilon_{2} x)]^{c},
\end{equation}
where $ \epsilon^2=(\epsilon_{1})^2=(\epsilon_{2})^2=1$ and $c$ is either complex conjugation, or Berezin conjugation or unity, we obtain
$T-$ symmetric, $P-$symmetric and $PT-$ symmetric  nonlocal integrable equations provided that the curvature and torsion tensors of the homogeneous space satisfy certain conditions. As an example to such soliton connections we have studied super AKNS system \cite{gur2},\cite{gur3} and we have recently shown that super integrable systems admit also Ablowitz-Musslimani type of nonlocal reductions. We have found integrable nonlocal super NLSE and nonlocal super mKdV equations \cite{gur4}.

\vspace{1cm}

\section{Acknowledgment}
  This work is partially supported by the Scientific
and Technological Research Council of Turkey (T\"{U}B\.{I}TAK).


\begin{thebibliography}{99}
\bibitem{ab1} Mark. J. Ablowitz and Ziad H. Musslimani, {\it Integrable Nonlocal Nonlinear Schrodinger Equation}, Phys.Rev.Lett.
{\bf 110}, 064105 (2013).

\bibitem{ab2} Mark. J. Ablowitz and Ziad H. Musslimani,{\it Integrable nonlocal nonlinear equations}, Studies in Applied Mathematics 00.1-53
(DOI:10.1111/sapm.12153). A special volume dedicated to Professor David J. Benney.

\bibitem{ab3} Mark. J. Ablowitz and Ziad H. Musslimani,{\it Inverse scattering transform for the integrable nonlocsl nonlinear Scrodinger
equation}, Nonlinearity {\bf 29}, 915-946 (2016).


\bibitem{fok} A. S. Fokas, {\it Integrable multidmensional versions of the nonlocal Schrodinger equation}, Nonlinearity, {\bf 29}, 319-324 (2016).

\bibitem{ma} Li-Yuan Ma, Shou-Feng Shen and Zuo-Nong Zhu, {\it Integrable nonlocal complex mKdV equation: soliton solution and gauge equivalence}, {\tt arXiv: 1612.06723 [nlin.SI].}

\bibitem{gerd} V.S. Gerdjikov and A. Saxena, {\it Complete integrability of Nonlocal Nonlinear Schrodinger Equation}, {\tt arXix: 1510.00480 [nlin.SI].}

\bibitem{sin} Debdeep Sinha and Pisush K. Ghosh, {\it Integrable nonlocal vector nonlinear Scgrodinger equation with self-induced parity-time symmetric potential}, Physics Letters {\bf A 381}, 124-128 (2017).

\bibitem{fk} Allan P. Fordy and Peter P. Kulish, {\it Nonlinear Scrodinger Equations and Simple Lie Algebras},
Commun. Math. Phys. {\bf 89}, 427-443 (1983).

\bibitem{for1} Allan P. Fordy , {\it Derivative nonlinear Schrodinger equations and Hermitian symmetric spaces}, J. Phys. A: Math. Gen. {\bf 17}, 1235-1245, (1984).

\bibitem{gerd1}  V. S. Gerdjikov, D. G. Grahovski and R. I. Ivanov, {\it On the integrable wave interactions and Lax pairs on symmetric spaces}, {\tt arXiv: 1607.06940 [nlin.SI]},in the special issiue on Mathematical modelling and physical dynamics of solitary waves: From continuum mechanics to field theory, Rds. Ivan C. Christov, M. D. Todorov, S. Yoshida. (Wave Motion,http://dx.doi.org/10.1016/j.wavemoti.2016.07.012)

\bibitem{gerd2}  V. S. Gerdjikov, D. G. Grahovski and R. I. Ivanov, {\it On the N-wave Equations with PT symmetry}, Theoretical and Mathematical Physics, {\bf 188}, No.3, 1305-1321 (2016).

\bibitem{gerd3} V. S. Gerdjikov, {\it On nonlocal models of Kulish-Sklyanin type and generailzed Fourirer transforms}, {\tt arXiv: 1703:03705}.

\bibitem{gur1} M. G{\" u}rses, {\" O}. O{\~ g}uz and S. Saliho{\~ g}lu, {\it Nonlinear partial differential equations on homogeneous spaces}, Int. J. Mod. Phys. A {\bf 5}, 1801-1817 (1990).

\bibitem{gur2}  M. Gurses and O. Oguz, {\it Super AKNS Scheme}, Phys. Lett.,{\bf A108}, 437 (1985).

\bibitem{gur3} M. Gurses and O. Oguz, {\it A Super Soliton Connection}, Letters in Mathematical Physics, {\bf 11}, 235-246 (1986).

\bibitem{gur4} M. Gurses, {\it Nonlocal Super Integrable Equations} (in progress).



\end{thebibliography}
\end{document}